  \documentstyle[twocolumn,prl,aps,psfig,epsf,amssymb,amsfonts]{revtex}
\begin{document}
%
%
\renewcommand{\Re}{\operatorname{Re}}
\renewcommand{\Im}{\operatorname{Im}}
\newcommand{\Tr}{\operatorname{Tr}}
\newcommand{\sign}{\operatorname{sign}}
\newcommand{\dd}{\text{d}}
\newcommand{\q}{\boldsymbol q}
\newcommand{\p}{\boldsymbol p}
\newcommand{\rr}{\boldsymbol r}
\newcommand{\pp}{p_v}
\newcommand{\vv}{\boldsymbol v}
\newcommand{\I}{{\rm i}}
\newcommand{\pphi}{\boldsymbol \phi}
\newcommand{\ds}{\displaystyle}
\newcommand{\be}{\begin{equation}}
\newcommand{\ee}{\end{equation}}
\newcommand{\bea}{\begin{eqnarray}}
\newcommand{\eea}{\end{eqnarray}}
\newcommand{\Acl}{{\cal A}}
\newcommand{\Rcl}{{\cal R}}
\newcommand{\Tcl}{{\cal T}}
\newcommand{\Tmin}{{T_{\rm min}}}
\newcommand{\Toff}{{\langle \delta T \rangle_{\rm off} }}
\newcommand{\Roff}{{\langle \delta R \rangle_{\rm off} }}
\newcommand{\RoffI}{{\langle \delta R_I \rangle_{\rm off} }}
\newcommand{\RoffII}{{\langle \delta R_{II} \rangle_{\rm off} }}
\newcommand{\dg}{{\langle \delta g \rangle_{\rm off} }}
\newcommand{\rd}{{\rm d}}
\newcommand{\br}{{\bf r}}
\newcommand{\la}{\langle}
\newcommand{\ra}{\rangle}

\twocolumn[\hsize\textwidth\columnwidth\hsize\csname @twocolumnfalse\endcsname
%
%
\draft

\title{Semiclassical Theory of Chaotic Quantum Transport}

\author{Klaus Richter$^{(1)}$ and Martin Sieber$^{(2)}$}

\address{
$^{(1)}$ Institut f\"ur Theoretische Physik, Universit\"at Regensburg, 
         93040 Regensburg, Germany}

\address{
$^{(2)}$ School of Mathematics, University of Bristol,
         University Walk, Bristol BS8 1TW, England}

\date{\today}

\maketitle

\begin{abstract}
We present a refined semiclassical approach to the
Landauer conductance and Kubo conductivity of clean chaotic mesoscopic systems.
We demonstrate for systems with uniformly hyperbolic dynamics
that including off-diagonal contributions to double sums over 
classical paths gives a weak-localization correction 
in quantitative agreement with results from random matrix theory.
We further discuss the magnetic field dependence.
This semiclassical treatment accounts for current conservation.
\end{abstract}

\pacs{03.65.Sq,73.20.Fz,05.45.Mt}

%
]



\narrowtext

Among the prominent wave phenomena which constitute 
mesoscopic behavior of small phase-coherent conductors, 
weak localization (WL) represents one key mechanism. This
quantum effect shows up as a decrease in the average conductivity
with respect to the classical one. WL, originally observed for
light and electron waves in disordered samples, has been extensively
studied during the last decade for electrons in ballistic conductors,
i.e.\ mesoscopic cavities or quantum dots where the 
elastic mean free path is considerably larger than the system size. 
WL is attributed to constructive interference of electron waves 
which are either coherently backscattered from impurities in disordered systems
or multiply reflected at the boundaries of ballistic devices. 

In describing ballistic transport semiclassical methods have attracted much 
interest, since they establish a direct link between quantum transport 
properties and features of the corresponding classical dynamics,
e.g.\ chaotic, integrable, or mixed behaviour
\cite{Jal00,Ric00}. This was demonstrated for clean microstructures
in a seminal semiclassical approach\cite{BJS93}
to the average reflection in the Landauer framework: 
the  WL peak profile (as a function of a magnetic field) was
shown to be Lorentzian for chaotic systems,
probing in an impressive way the imprint of the classical dynamics 
on a measured quantum effect\cite{CBPW94}. However, while the 
line shape of the WL peak agreed with results from 
random matrix theory (RMT), the approach 
turned out to be inadequate to give the correct WL magnitude for 
ballistic systems, contrary to the diffusive case\cite{CS86}.
This long-lasting problem to semiclassically obtain the correct
leading-order quantum correction to the conductance
is related to the so-called {\em diago\-nal approximation} used.
It is based on the consideration of a restricted class of
pairs of paths built from each backscattered orbit and
its time-reversed partner,
which moreover violates current conservation\cite{note1}. 
Similarly, WL is not captured in a related semiclassical approach
to the Kubo conductivity of ballistic systems\cite{OH95}.

The possible relevance of pairs of non-identical backscattered paths,
differing slightly in their initial directions, was first pointed 
out by Argaman \cite{Arg95}. Later, Aleiner and Larkin\cite{AL96} 
approached the problem of ballistic WL using both, perturbation 
theory and supersymmetrical methods to derive a RMT result for the
conductance. However, their techniques still
rely on the presence of quantum scatterers 
(to regularize the Liouville operator) and strictly speaking
do not treat the case of a clean, disorder free, system.
Their approach was semiclassically interpreted in Ref.~\cite{TN97} 
arguing that diffraction-induced small angle scattering is relevant for 
ballistic WL.

Here we present an adequate, current-conserving
semiclassical treatment 
of the problem to quantitatively describe the average 
quantum conductance in {\em clean} chaotic systems without relying on
any diffraction or impurity scattering effects.
We consider the leading-order off-diagonal contribution in a
semiclassical loop expansion of the Landauer conductance.
The relevant off-diagonal terms consist of pairs of
orbits which are very close almost everywhere (in configuration
space), and differ only in whether they undergo or avoid
a self-intersection with small crossing angle\cite{note2}. 
Our results are strictly derived for chaotic systems
with uniformly hyperbolic dynamics,
but related results for ballistic cavities indicate\cite{TR01} 
that they apply to general chaotic systems.

We first compute semiclassical conductance contributions beyond
the diagonal approximation in the Landauer framework and later 
return to the corresponding problem in the Kubo formalism.
Consider a two-dimensional, classically chaotic clean cavity with two leads
of width $w (w')$ attached that support $N (N')$ open
channels. The Landauer formula for the conductance $G$ then reads\cite{FL81}
\begin{equation}
\label{eq:Landauer}
G(E,B) = 2\frac{e^2}{h}  \ \sum_{n=1}^{N'}  \sum_{m=1}^N |t_{nm}(E,B)|^2  \; .
\end{equation}
Here $t_{nm}(E,B)$ denotes the 
transition amplitude between incoming and outgoing channels $m$ and $n$
at energy $E$ in the presence of a magnetic
field $B$. We first consider the case of time-reversal symmetry, 
$B\!=\!0$, and return to the $B$-dependence of WL later.
We assume that the ergodic time is much 
smaller than the escape time $\tau$ of the cavity and that 
contributions from direct, lead-connecting processes are negligible. 
Then the following RMT results for the transmission and reflection 
amplitudes hold which we give for later reference \cite{RMT,Bee97}:
\bea
\label{eq:rmt}
|t_{nm}|^2  & = &  \frac{1}{N\! +\! N'\! + \! 1}  
   =  \frac{1}{N + N'}   
   \sum_{k=0}^\infty \left[\frac{-1}{N + N'}\right]^k \; , \\
\label{eq:rmt2}
|r_{nm}|^2 & = &   \frac{1 + \delta_{nm}}{N\! +\! N'\! + \! 1}
   =  \frac{1+ \delta_{nm}}{N + N'}  \
  \sum_{k=0}^\infty \left[\frac{-1}{N + N'}\right]^k \; .
\eea

\begin{figure} 
\begin{center}
\mbox{ \psfig{figure=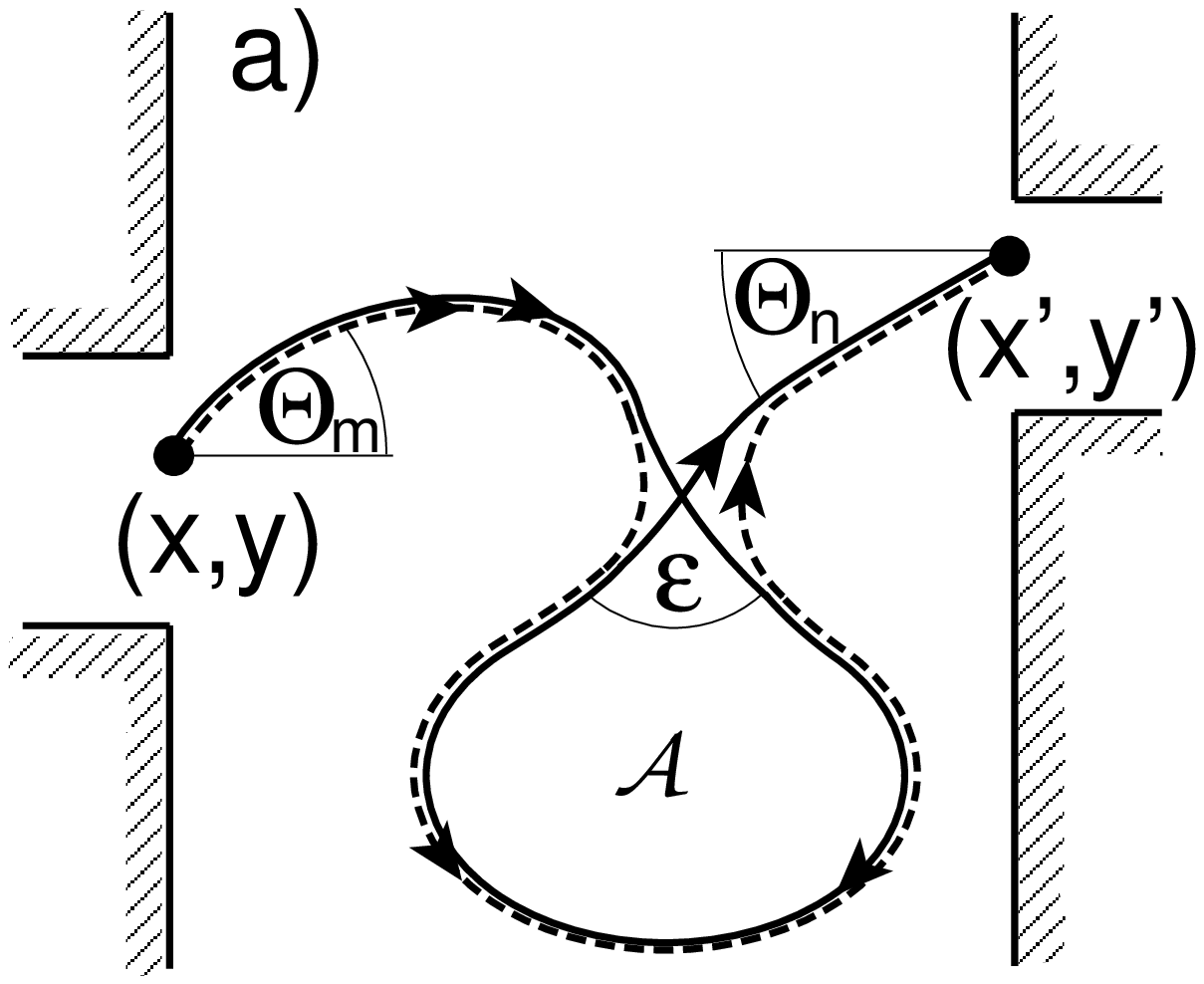,width=0.2\textwidth,angle=0}
 \hspace*{10mm} 
\psfig{figure=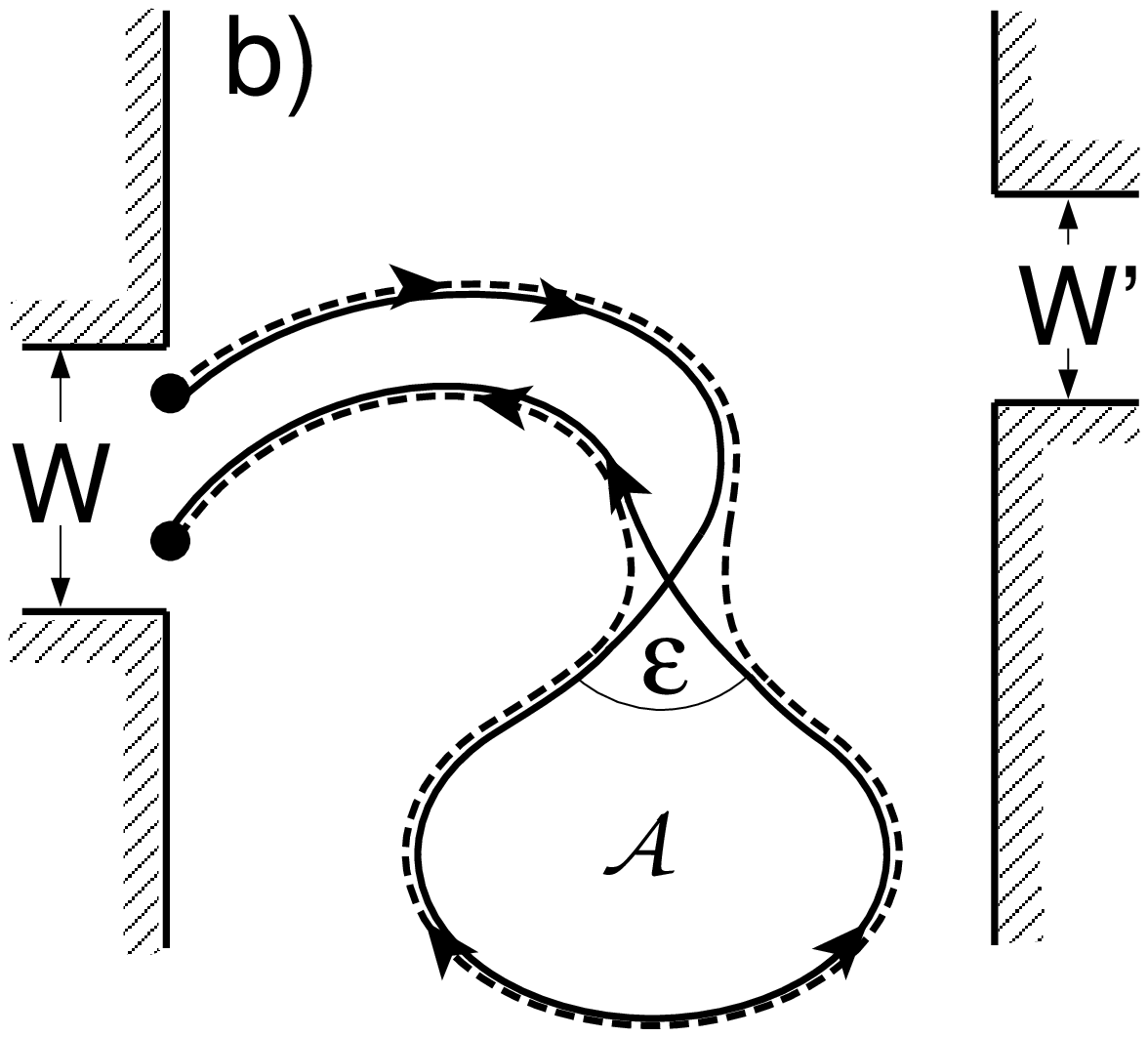,width=0.20\textwidth,angle=0} }
 \end{center}
\caption{
Sketch of an off-diagonal trajectory pair formed by 
a self-intersecting classical orbit (solid line)
with small crossing angle $\varepsilon$ and a neighboring orbit 
(dashed) differing mainly in the region around the self-intersection.
The paths represent orbits with many reflections at 
the system boundaries.
They con\-tri\-bute to the quantum transmission (a) and reflection (b).
}
\label{fig:orbits}
\end{figure}

Our  conductance calculation is based on
the semiclassical representation of transmission amplitudes\cite{note3},
\begin{equation}
\label{eq:sc-tnm}
t_{nm} \simeq -\sqrt{\frac{\pi\hbar}{2 w w'}} \!
 \sum_{\gamma(\bar{n},\bar{m})} 
 \frac{\Phi_\gamma \exp[(\I/\hbar)S_\gamma] }{ 
| \cos \theta'_{\bar{n}} \cos \theta_{\bar{m}}
   M^\gamma_{21}|^{1/2}} \; .
\end{equation}
The sum runs over all lead-connecting trajectories $\gamma$ which 
enter into the cavity at $(x,y)$ with an angle
$\sin\Theta_{\bar{m}} = \bar{m} \pi /(k w)$ and exit the cavity at $(x',y')$
with angle $\sin\Theta_{\bar{n}} = \bar{n} \pi /(k w')$, where $\bar{n} = \pm n$,
and $p = \hbar k$ is the momentum, see Fig.~\ref{fig:orbits}(a).
In Eq.~(\ref{eq:sc-tnm}), $S_\gamma$ is the classical 
action,  $M^\gamma_{21}$ an element of the stability matrix, and
$\Phi_\gamma = {\rm sgn}(\bar{n}) {\rm sgn}(\bar{m})
\exp[i\pi(\bar{m}y/w\!-\!\bar{n}y'/w' \!-\! \mu_\gamma/2 \!+\! 1/4)]$ is
a phase factor where $\mu_\gamma$ contains the Morse index.
An expression corresponding to Eq.~(\ref{eq:sc-tnm}) holds true for $r_{nm}$
in terms of paths reflected back.

The Landauer Eq.~(\ref{eq:Landauer}) contains 
products $t_{nm} t_{nm}^\ast$ which semiclassically amounts to evaluate 
double sums over an infinite number of trajectory pairs. 
In a treatment of the energy-averaged conductance
most pairs, consisting of orbits with uncorrelated actions,
will cancel each other upon summation. The existing
  semiclassical approach\cite{BJS93} is based on 
the diagonal approximation, where only pairs of identical
orbits or orbits related to each other by time inversion
are taken into account. Then the phase factors from
Eq.~(\ref{eq:sc-tnm}) cancel, and one has 
$ |t_{nm}|_{\rm diag}^2 = \pi \hbar/(2w w') 
 \sum_{\gamma(\bar{n},\bar{m})} 
  | \cos {\theta'}_{\bar{n}} \cos \theta_{\bar{m}} M^\gamma_{21}|^{-1}$.

First we give an alternative further evaluation of this 
expression employing the sum rule\cite{RS02,note4}
\begin{equation}
\label{eq:sum-rule}
\sum_{\gamma(y',{\theta}_n';y,\theta_m)} \! \!
\frac{\delta(T-T_\gamma)}{|M^\gamma_{21}|}
\simeq \frac{\cos\theta'_n\cos\theta_m}{\Sigma(E)}
 {\rm d} y \, {\rm d} y' \ \rho(T) \; .
\end{equation}
The sum runs over all orbits with periods $T_\gamma$,
which begin and end in intervals  
${\rm d} y'$ and ${\rm d} y$ around  $y'$ and $y$
with fixed orientations of the initial and final velocities.
$\Sigma(E)$ is the energy surface in phase space;
$\Sigma(E) = 2\pi mA$ for billiards of area $A$.  
The factor $\rho(T) \sim \exp(-T/\tau)$ (for $T \!\rightarrow \!\infty$)
accounts for the exponential loss of particles with velocity 
$v$ which escape through the openings characterized by the 
 escape rate
\begin{equation}
\label{eq:escape-rate}
\frac{1}{\tau} =  \frac{v(w+w')}{A\pi} = \frac{\hbar}{mA}(N+N') \; .
\end{equation}
Upon applying the sum rule (\ref{eq:sum-rule}) to the diagonal
contribution, integrating over the lead cross sections,
and including a factor 4 for each tupel 
$(\bar{n},\bar{m})$ one finds for the transmission coefficient of
an ergodic system
\begin{equation}
\label{eq:tnm-diag}
|t_{nm}|_{\rm diag}^2 =  
4 \frac{\pi \hbar/2}{2\pi m A} \int {\rm d} T \  e^{-T/\tau}  
            =  \frac{1}{N + N'} \; .
\end{equation}

Correspondingly, the quantum reflection coefficient reads in the diagonal
approximation
\begin{equation}
\label{eq:rnm-diag}
|r_{nm}|_{\rm diag}^2 =  \frac{1}{N + N'} + \frac{\delta_{nm}}{N + N'} \; .
\end{equation}
The semiclassical evaluation at this level yields the
$(k\!=\!0)$-term of the RMT result (\ref{eq:rmt},\ref{eq:rmt2}). 
Note that the use of the sum rule (\ref{eq:sum-rule}) allows us to
compute {\em individual} transmission and reflection coefficients, while 
Ref.\ \cite{BJS93} gives results only for the entire classical transmission
and reflection. 

Summing the first term in Eq.~(\ref{eq:rnm-diag}) over all channels 
yields the classical reflection $R = N^2/(N+N')$. 
The se\-cond term in Eq.~(\ref{eq:rnm-diag})
arises from contributions to $|r_{nn}|^2$ from backscattered orbits paired  
with their time-reversed partners (elastic enhancement). This gives rise 
to the di\-a\-go\-nal contribution to WL, $\delta R_{\rm diag} = N/(N+N')$,
as derived in \cite{BJS93}. 
In the limit $N\!=\!N' \rightarrow
\infty$ one has  $\delta R_{\rm diag} = 1/2$,
deviating from the  RMT result $\delta R_{\rm RMT} = 1/4$. 


In the following we go beyond the diagonal approximation and
consider pairs of different trajectories as sketched in Fig.~\ref{fig:orbits}
in coordinate space. They consist of a long, self-intersecting orbit 
(solid line in Fig.~\ref{fig:orbits}a),b)) with small crossing angle $\varepsilon$
forming a closed loop and a second orbit in its close vicinity (dashed line).  
The two orbits follow the two open trajectory segments, beginning and ending at 
(exponentially) close points
at the lead mouths, in the same but the loop in opposite direction. 
Given a self-intersecting orbit with small $\varepsilon$ we showed that
the neighbouring orbit indeed exists by linearizing the motion in its vicinity
\cite{RS02}.
The action difference $\Delta S(\varepsilon)$ of the two orbits is computed by expanding the
action around the self-intersecting orbit up to second order. The 
resulting formula for  $\Delta S$, expressed through the elements of
the stability matrices for the loop and the two open segments, is 
rather involved. Hence, we will focus from now on onto chaotic systems with 
uniformly hyperbolic dynamics characterized by a single Lyapunov 
exponent $\lambda$ and no conjugate points. One then finds\cite{RS02}
\begin{equation}
\label{eq:actdiff}
\Delta S(\varepsilon)  
 \approx  \frac{p^2 \varepsilon^2}{2 m \lambda}  \; .
\end{equation}
Since a partner orbit
is associated with each self-intersection with small crossing angle,
we compute the conductance contribution from all such orbit pairs by first
summing for each orbit $\gamma(\bar{n},\bar{m})$ 
over all  $\varepsilon$ self-intersections 
and finally by integrating over $\varepsilon$.
Using Eq.~(\ref{eq:actdiff}) for the action differences 
occuring in $t_{nm} t_{nm}^\ast$ (with $t_{nm}$ from
Eq.~(\ref{eq:sc-tnm})) and 
the sum rule
(\ref{eq:sum-rule}) one finds for this `loop' contribution
\bea
|t_{nm}|_{\rm loop}^2 &  \simeq & \frac{\pi \hbar}{w w'}
 \sum_{\gamma(\bar{n},\bar{m})}  
  \frac{\delta(T-T_\gamma)}{
  | \cos {\theta'}_{\bar{n}} \cos \theta_{\bar{m}} M^\gamma_{21}|} 
  I(\varepsilon,T) \nonumber \\
& \simeq & 
\label{eq:tnm-int}
\frac{2 \hbar}{m A} \int {\rm d} T  e^{-T/\tau} I(\varepsilon,T)
\eea
with 
\begin{equation}
\label{eq:Ieps}
I(\varepsilon,T) = 
  {\rm Re} \int_0^\pi {\rm d} \varepsilon
 P(\varepsilon, T)  
 \exp \left( \frac{i p^2 \varepsilon^2}{2 \hbar m \lambda} \right) \; .
\end{equation}
In the semiclassical limit ($\hbar \rightarrow 0$) the contribution 
from small angles is dominant. In Eq.~(\ref{eq:Ieps}),
the density $P(\varepsilon, T)$ of self-crossings with angle 
$\varepsilon$ for a long orbit of time $T$ can be expressed as
an integral over all loops, associated with the self-crossings,
with times 
$T_{\rm min}(\varepsilon) < t < T$:
 \begin{equation}
 \label{eq:P-integral}
 P(\varepsilon,T)  \simeq  2 m v^2 
\int_{\Tmin(\varepsilon)}^T
\dd t  (T-t) \sin (\varepsilon) \, p_{\rm erg} \; ,
 \end{equation}
where $p_{\rm erg} = 1/(2\pi m A)$ is
the ergodic classical return probability.
The lower cutoff accounts for the fact that, owing to the exponential
divergence of neighboring orbits in a
 hyperbolic system, a minimum time $\Tmin(\varepsilon)$ is
required to form a closed loop from two trajectories starting
at the crossing with initial angular difference $\varepsilon$.
Hence $\Tmin(\varepsilon)$ can be estimated from
$c \simeq \varepsilon \exp(\lambda \Tmin(\varepsilon)/2)$ with $c$
of order  $\pi$.
Detailed numerical and analytical studies \cite{SR01} 
have shown that this indeed holds true and that the number of
crossings  
for $ T \rightarrow \infty$ is given by 
\begin{equation}
\label{eq:Pvont}
P(\varepsilon,T) \, \dd \varepsilon \sim
\frac{T^2  v^2 }{\pi A}
\frac{\sin \varepsilon}{2}
\left[1 - 2\frac{\Tmin(\varepsilon)}{T} \right]
\, \dd \varepsilon \; 
\end{equation}
with  $\Tmin(\varepsilon) = -(2/\lambda) \log(\varepsilon/c)$.
The integral (\ref{eq:Ieps}) over the leading-order $T^2$ term in 
Eq.~(\ref{eq:Pvont}) is purely imaginary, and thus its contribution vanishes. 
However, the contribution to $I(\varepsilon,T)$ of the second, logarithmic 
term in Eq.~(\ref{eq:Pvont}) 
is finite and gives $-(\hbar/2mA)T$, independent of $\lambda$.
We then obtain from Eq.~(\ref{eq:tnm-int})
\begin{equation}
\label{eq:tnm-loop}
|t_{nm}|_{\rm loop}^2 \simeq
- \left[\frac{\hbar}{m A}\right]^2 \! \int {\rm d} T T  e^{-T/\tau} 
=  \frac{-1}{(N + N')^2} \; .
\end{equation}
Hence, the lack of short loops with $t < \Tmin(\varepsilon)$
gives rise to a negative quantum correction to the transmission.

Correspondingly, we find for the loop correction to the
reflection coefficient
\begin{equation}
\label{eq:rnm-loop}
|r_{nm}|_{\rm loop}^2 
=  -\frac{1+ \delta_{nm}}{(N + N')^2} \; .
\end{equation}
Here, as for the diagonal contribution (\ref{eq:rnm-diag}),
backscattering into the same channel is twice as probable. 

Summing over all initial and final channels
we obtain for the leading-order quantum transmission
and reflection $\delta T_{\rm loop} = -N N'/(N + N')^2 $ and
 $\delta R_{\rm loop} = -N (N+1)/(N + N')^2 $. For
$N,N' \gg 1$ we have $\delta R_{\rm diag} + \delta R_{\rm loop} \simeq
  NN'/(N + N')^2 = -\delta T_{\rm loop}$. This implies
conservation of the average current 
in the semiclassical limit. 
{\em Considering off-diagonal terms
allows us to semiclassically compute WL corrections consistently either in
transmission or reflection}. They precisely coincide with the RMT result
$\delta T_{\rm RMT} = -1/4$  for
$N\!=\!N'\rightarrow \infty$. Comparison with the RMT results
for finite $N, N'$ suggests that the $k$th order terms in 
Eqs.~(\ref{eq:rmt},\ref{eq:rmt2}) correspond to semiclassical 
$k$-loop contributions; the diagonal terms are considered
as 0-loop and the orbits in Fig.~\ref{fig:orbits} as 1-loop terms.

Since the closed loops formed by the off-diagonal orbit pairs are traversed
in opposite directions, see Fig.~\ref{fig:orbits}, these orbits acquire an 
additional action or phase difference in the presence of a weak 
magnetic field $B$ due to the flux enclosed. For a uniform perpendicular 
field the action difference is $4\pi \Acl B/\phi_0 $, where $\Acl$ is 
the area of the loops and $\phi_0$ the flux quantum. 
We assume that the distribution 
$p(t;\Acl)$ of enclosed areas for trajectories of 
time $t$ is Gaussian with a  system specific parameter $\beta$,
\begin{equation}
\label{eq:area-dist}
p(t;\Acl) \simeq \frac{1}{\sqrt{2\pi t \beta}}
\ \exp\left(-\frac{\Acl^2}{2t\beta}\right) \; .
\end{equation}
This is usually well fulfilled for chaotic systems\cite{Jal00,Ric00,BJS93}.
For finite $B$-fields we have to perform an additional integration
of the field-induced phase differences over the area distribution:
$ \int_{-\infty}^\infty  \dd \Acl  p(t;\Acl) \cos(4\pi \Acl B/\phi_0)
=  \exp (-t/t_B) $, with the magnetic time
$
t_B\! = \! (4\pi \beta B/\phi_0)^{-2}. 
$
Up to timescales  $\Tmin (\varepsilon)$ a negligible flux is
enclosed by loops with small crossing angles. 
We consider this by a respective time shift 
when inserting $\exp (-t/t_B) $ into the integral 
(\ref{eq:P-integral}) over loop lengths:
\bea
\label{eq:PofB-integral}
 & P_B(\varepsilon,T) & \\
&  \simeq  & \!\!  
  \frac{v^2}{\pi A} \sin \varepsilon
\int_{\Tmin(\varepsilon)}^T \dd t  (T-t) \ e^{-[t-\Tmin(\varepsilon)]/t_B} 
\nonumber \\
& \sim & \!\! \frac{v^2 t_B^2}{\pi A}  \sin \varepsilon
  \left[ \frac{T}{t_B}\! -\! 1 \! + \! \frac{\Tmin(\varepsilon)}{t_B}
 (e^{-T/t_B}\!-\! 1) + \ldots \right] . 
\nonumber 
\eea
In Eq.~(\ref{eq:PofB-integral}) we used
$\Tmin(\varepsilon) \ll t_B$. This corresponds to the original
assumption, $\Tmin(\varepsilon) \ll \tau$, in the range of interest,
$\tau \sim t_B$. Only the term linear in $\Tmin(\varepsilon)$ 
contributes to the integral (\ref{eq:Ieps}), and we eventually
obtain, after computing the $T$-integral (\ref{eq:tnm-int}),
a Lorentzian field dependence of the transmission coefficient:
$|t_{nm}(B)|_{\rm loop}^2 \simeq
|t_{nm}(0)|_{\rm loop}^2 / (1 + \tau/t_B)$, 
A corresponding result applies to $|r_{nm}(B)|_{\rm loop}^2$.
This coincides with the Lorentzian WL lineshape obtained in the
diagonal approximation\cite{BJS93}, making clear why the diagonal 
terms already qualitatively account for the WL peak profile.
The entire WL correction from the diagonal and offdiagonal 
(1-loop) contribution
then reads, in terms of the classical reflection and transmission 
coefficients $r_{\rm cl}$ and $t_{\rm cl}$,
\begin{equation}
\label{eq:WLofB}
\delta R(B) \simeq  \frac{t_{\rm cl} r_{\rm cl}}{1 + \tau/t_B} \; .
\end{equation}
Our refined semiclassical approach to the 
Landauer conductance yields the correct WL magnitude {\em and} lineshape.

The {\em Kubo conductivity} reads, in terms of advanced (retarded)
Green functions $G^\pm({\bf r, r'};E) $
(for a system of area $A$),
$
\sigma = -[e^2 \hbar/ (4\pi A)] {\rm Tr}\{ \hat{v}_x \Delta G
               \hat{v_x} \Delta G \}, 
$
with $ \Delta G = G^+ - G^-$. 
The trace is semiclassically evaluated in position representation
by approximating the products of
Green functions involved through double sums over classical
phase-carrying paths. Pairing identical orbits in
the diagonal approximation leads to the classical Kubo conductivity
\cite{Ric00,OH95,Arg95}; off-diagonal terms are again required 
to compute WL for chaotic systems.
As a prototype of an extended clean chaotic system
consider, e.g., a two-dimensional Lorentz gas. This has been
experimentally 
realized by regular arrays or disordered ensembles of antidots in 
two-dimensional semiconductor heterostructures\cite{YLWR00}.
The antidots act as classical scatterers giving rise to 
diffusive motion on long time scales, while the dynamics
for intermediate times is governed by chaotic scattering.

Our semiclassical treatment of WL is based on 
off-diagonal pairs of paths which 
have much in common with the orbits discussed above (Fig.~1(b)):
they consist of one long self-intersecting trajectory
being backscattered after multiple bounces with antidots
with nearly opposite momentum and a neighboring orbit
which follows the loop formed by the first in opposite
direction. A careful treatment of the 
conductivity trace integral
for such pairs of backscattered paths (involving again
cutoff times logarithmic in the crossing angle)
gives rise to a non-vanishing 
negative quantum contribution $\delta \sigma$ at $B\!=\!0$\cite{RS02}. 

This WL correction for chaotic  systems
with classical scatterers 
turns out to coincide with that from disordered 
systems with quantum impurity scattering.
We find
\begin{equation}
\label{eq:WL-Kubo}
 \delta \sigma \simeq 
  -(e^2/\pi h)  
   \ln(t_\phi/t_{\rm el}) \; ,
\end{equation}
where $t_\phi$ is the phase-coherence time and $t_{\rm el}$ 
the elastic scattering time due to reflections at the
antidots. Diffusive motion on long time scales, 
accounted for in a sum rule similar to Eq.~(\ref{eq:sum-rule}),
is reflected in the $\ln$.
Eq.\ (\ref{eq:WL-Kubo}) coincides with the 
result of Ref.~\cite{AL96} for antidot systems when
$t_{\rm \phi}$ is large compared to the
Ehrenfest time. 

The approach above can be generalized to  treat 
linear-response functions of other observables\cite{RS02}.

To conclude, a semiclassical treatment 
beyond the diagonal approximation is appropriate to compute
quantum corrections to the average conductance
in clean chaotic conductors, 
both in the Landauer and Kubo framework. Chaotic classical
dynamics is responsible for a logarithmic angular dependence of
the classical return probability, respectively the loops involved,
which turns out to be crucial for computing weak localization.
Numerical results for billiards\cite{TR01} 
show that this $\log\varepsilon$-dependence 
holds true also for nonuniformly hyperbolic systems indicating 
that the mechanism presented here is rather general.
A semiclassical evaluation of higher-order loop corrections
is yet to be performed. While such 
terms are not negligible for the
spectral form factor, the one-loop corrections considered here
play the dominant r\^ole for quantum transport in the
mesoscopic regime.

This work was supported by the
Deutsche Forschungsgemeinschaft under contract Ri-681/5-1.




\begin{thebibliography}{10}
\bibitem{note1} 
As a severe shortcoming of the diagonal approximation the magnitude of 
the WL peak depends on the way it is semiclassically derived
and may even vanish\cite{Ric00}.
\bibitem{note2} 
	Analogous pairs of periodic orbits have recently been used 
	to derive the $\tau^2$-term in the spectral form factor of 
	RMT\cite{SR01}. These orbit pairs are ballistic analogues of 
	corresponding objects in the diffusive 
	regime\cite{WLS99}.
\bibitem{note3} 
It is obtained from a semiclassical approximation of the 
Green functions in the Fisher-Lee relation\cite{FL81};
see \cite{BJS93}.
\end{thebibliography}
\end{document}